\documentclass[twocolumn
]{revtex4-2}

\usepackage{amsmath}
\usepackage{caption}
\usepackage{graphicx}
\usepackage{latexsym}
\usepackage{dcolumn}
\usepackage{hyperref}
\usepackage{comment}
\usepackage{chemformula}
\usepackage{bm}
\usepackage{xcolor}
\usepackage{array}
\usepackage{enumitem}

\begin{document}




\title{Machine Learning-Driven Chemical Reactor Network Modeling of the Sandia-D Flame}

\author{Nicolas J. Tricard}
\thanks{Equal contribution}
\affiliation{Department of Mechanical Engineering, Massachusetts Institute of Technology, Cambridge, MA 02139, USA}

\author{Benjamin C. Koenig}
\thanks{Equal contribution}
\affiliation{Department of Mechanical Engineering, Massachusetts Institute of Technology, Cambridge, MA 02139, USA}

\author{Sili Deng}
\thanks{Corresponding author: silideng@mit.edu}
\affiliation{Department of Mechanical Engineering, Massachusetts Institute of Technology, Cambridge, MA 02139, USA}


\begin{abstract}
Turbulent combustion simulations are crucial for many scientific and engineering systems. However, the high cost to fully resolve the complex multiscale and multiphysics behavior makes direct simulation typically infeasible. The equivalent reactor network (ERN) approach attempts to improve computational efficiency by replacing a multidimensional turbulent simulation with a series of much cheaper 0-D and 1-D chemical reactors, providing a surrogate model that retains detailed chemistry at the cost of simplified flow physics. However, their development remains a challenge, often requiring expert analysis, or automated approaches that sacrifice accuracy. 
In this work, we develop an automated machine-learning-assisted framework for constructing ERNs of the Sandia-D turbulent methane/air flame. Principal component analysis is first used to reduce high-dimensional thermochemical computational fluid dynamics (CFD) data to a low-dimensional latent space, where k-means clustering identifies physically interpretable flame regions used to initialize a reactor-network graph. This initialization is then refined using finite-difference gradient descent wrapped around non-differentiable Cantera reactor simulations. Across 30 Reynolds-averaged Navier-Stokes (RANS) simulations spanning a range of pilot temperatures and inlet methane compositions, the optimized 7-reactor ERN achieves a maximum-temperature $R^2$ score of 0.7945 while preserving a $\sim6000\times$ speedup over the CFD solver. Outlet CO prediction remains more challenging, with a final $R^2$ score of $-0.4183$, but improves substantially from the unoptimized clustering initialization. These results show that unsupervised thermochemical feature extraction can provide effective physics-informed initializations for ERN construction, while gradient-based refinement can significantly improve predictive accuracy without manual reactor-network design.
\end{abstract}

\maketitle


\section{Introduction\label{sec:introduction}} 

Large-scale turbulent combustion models are important for understanding and designing scientific and industrial processes \cite{echekki2010turbulent}. Their high cost often prevents the use of detailed simulations, which is a persistent challenge in the field but especially notable in applications requiring high sample counts, like sensitivity analysis, uncertainty quantification \cite{koenig2023kinetic}, or experimental design contexts \cite{chen2024fast}.

Many frameworks exist to simplify direct numerical simulations (DNS), such as the ubiquitous large eddy (LES) and Reynolds-averaged (RANS) simulations \cite{pitsch2006large}. LES simulations resolve the larger and cheaper turbulent effects while replacing the small, near-dissipation scales with simplified models. The even cheaper RANS approach models all scales. The equivalent reactor network (ERN) approach \cite{du2014equivalent, du2015equivalent} offers a more aggressive speedup by completely removing multidimensional flow effects altogether. ERNs model multidimensional turbulent reactors as interconnected networks of much cheaper canonical 0-D and 1-D reactors. These networks of reactors are known as chemical reactor networks (CRN), or ERN when directly applied as a substitute for CFD in a given real system.

These ERNs provide huge computational speedups and greatly facilitate applications that call for repeated sampling, but are challenging to construct and do not always provide strong or even acceptable accuracy. The series and parallel connections between these reactors, as well as the types, sizes, and residence times in each reactor, must be chosen carefully to accurately replicate the real-world combustion behavior across a range of inlet conditions. Expert-driven analysis and ERN construction offer bespoke networks that can capture complex chemical phenomena well \cite{du2014equivalent, du2015equivalent}. 
Numerous reports have demonstrated accurate ERNs for predicting pollutant emissions~\cite{Novosselov2006,Kanniche2010,Lee2011,Perpignan2018}, regulating combustor operating conditions~\cite{KALURI2018797}, or even reconstructing full flow-fields~\cite{Falcitelli01112002,SAVARESE2024105536}.
While these approaches have proven useful, we remark the scale of effort: skilled expert analysis is often required derive and verify a single high-accuracy ERN.

Automated approaches also exist. For example, recent work in 2025 automated the generation of ERNs of varying sizes to enable sensitivity analysis without requiring expert input, delivering sparse networks that included as few as 5 reactors and denser, high-accuracy networks with as many as 1250 reactors \cite{MODARRES2025100893}. 
Saverese \textit{et al.}~\cite{SAVARESE2023127945} applied an unsupervised k-means clustering algorithm to partition their CFD domain into regions based on temperature and reaction progress variable, and
Laborda \textit{et al.}~\cite{LABORDA2025108891} applied a similar k-means approach to bioreactor network arrangements. 
While these works begin to automate ERN construction, they rely largely on classical clustering and physics-based heuristics, leaving more modern machine learning approaches largely untapped in this context. Reactor counts required for accuracy thus remain high, with modest speedups over conventional CFD (only 4$\times$
in Ref.~\cite{MODARRES2025100893}).

\begin{figure}
\centering
\includegraphics[width=0.7\linewidth]{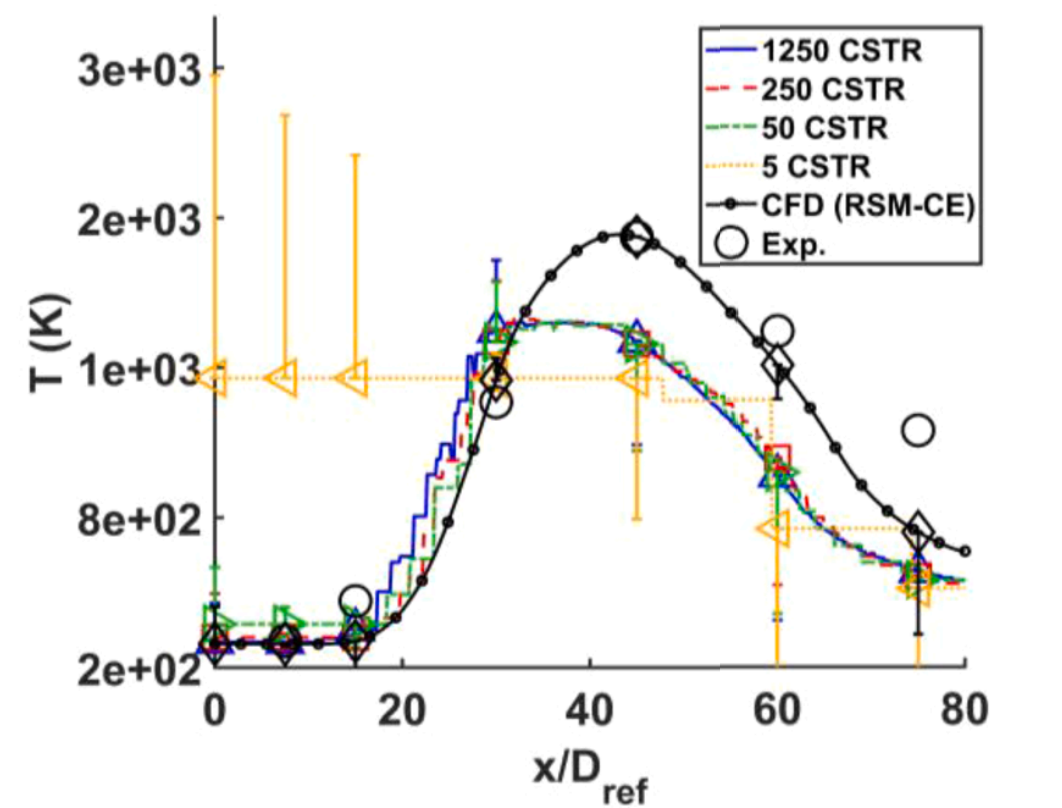}
	\caption{Traditional ERN temperature reconstruction metrics, reproduced from \cite{MODARRES2025100893}. The automatically generated 5-reactor ERN fails to reproduce the data.}
	\label{fig:Modarres}
\end{figure}

Our work aims to improve automated ERN development by applying recent machine learning tools to achieve higher accuracy with lower reactor counts and thus lower computational cost. We propose a combination of automatic dimension reduction and clustering algorithms paired with gradient descent to facilitate automatic ERN generation. Through this combination, we deliver a framework that offers computational cost reduction in forward evaluations while maintaining reasonable accuracy and without requiring detailed expert analysis. 
We apply this framework to the Sandia Flame D flame \cite{BARLOW19981087}, attempting to match integrated CO and peak temperature values based on a dataset of 30 RANS simulations.
This configuration is similar to that of the automatic sensitivity analysis work of \cite{MODARRES2025100893}. While our setup and workflow differ, we aim to deliver reasonable results that compare favorably against those of Ref. \cite{MODARRES2025100893}, especially in the low reactor count ERNs where the most computational speedup is observed. Those benchmark results are reproduced in Fig. \ref{fig:Modarres}.

\section{Methodology \label{sec:Methods}} 

\subsection{Chemical reactor models}

Given a complete thermochemical state (temperature, pressure, and species concentrations) $\mathbf{u}$, we can evolve a 0-D system forward in time as 
\begin{equation} \label{eq:kinetic_laws}
    \frac{d\mathbf{u}}{dt} = \mathbf{f}\left(\mathbf{u} \right),
\end{equation}
where $\mathbf{f}$ is the known chemical model (a high-dimensional collection of thermodynamic constants and kinetic rate laws). Here, we use the GRI-Mech 3.0 \cite{GRImech3}, which includes 53 species and 325 reactions. Full chemical equations are not reported here for brevity.

\begin{figure}[!htbp]
\centering
\includegraphics[width=0.95\linewidth]{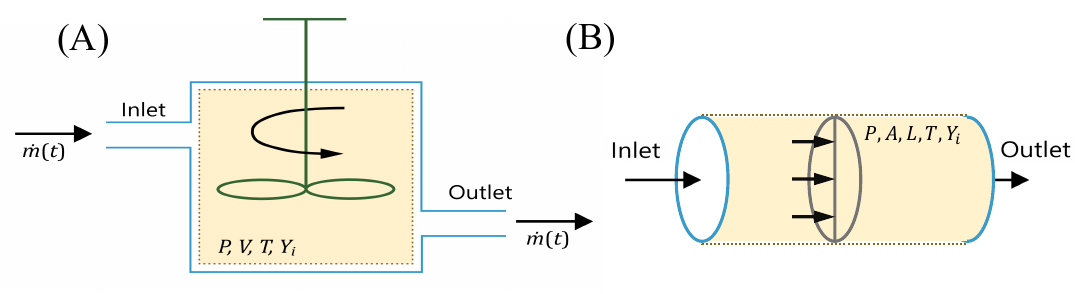}
	\caption{(A) WSR reactor visualization, adapted from \cite{cantera}. (B) PFR reactor visualization.}
	\label{fig:reactors}
\end{figure}

We model a turbulent flame with two simplified types of reactors, outlined in Table~\ref{tab:reactor_elements}. Full equations are available in Cantera \cite{cantera}. The well-stirred reactor (WSR) is a fixed-volume chamber with ideal mixing, modeled as a 0-D lumped mass (no spatial effects). A thermochemical composition and mass flow rate are provided at the inlet. To conserve mass, the outlet has the same mass flow rate, however the thermochemical composition will reflect the internal time-integrated evolution. The residence time can be represented directly as a function of the volume and mass flow rate: longer residence times run reactions toward completion, while short residence times have more transient effects in the outflow. WSRs are of use in ERNs when modeling mixing and recirculation zones, which see strongly turbulent rotating internal flow qualitatively matching the WSR model well. A schematic is in Fig. \ref{fig:reactors}(A).

The plug flow reactor (PFR) assumes no mixing and instead models the chemical evolution of individual parcels of a fluid as they move through a 1-D reactor. ``Plug flow'' implies no shearing or diffusion effects, so each axial slice of the reactor does not interact with adjacent slices. The PFR is 0-D in time and integrated in 1-D in space (vs. the 0-D in space WSR integrated 1-D in time). Similarly, the WSR models ideal mixing with zero transport, while the PFR models 1-D convection only with no mixing. PFRs are of use in larger ERNs when modeling strongly 1-D flows, such as in pipes, packed beds, or near inlets or nozzles. A schematic is in Fig. \ref{fig:reactors}(B).

\begin{table*}[t]
\caption{Reactor elements used in the ERN.}
\label{tab:reactor_elements}
\begin{ruledtabular}
\begin{tabular}{lll}
\textbf{Element} & \textbf{Idealization} & \textbf{Role in this work} \\
Well-stirred reactor (WSR) & Perfectly mixed 0-D reactor & Mixing, recirculation, distributed reaction \\
Plug flow reactor (PFR) & 1-D advective reactor & Jet-like or downstream reacting flow progression \\
Mixer/splitter & Algebraic mass redistribution & Connects reactor outputs and flow partitions \\
\end{tabular}
\end{ruledtabular}
\end{table*}

\begin{figure*}[!htbp]
\centering
\includegraphics[width=1.0\linewidth]{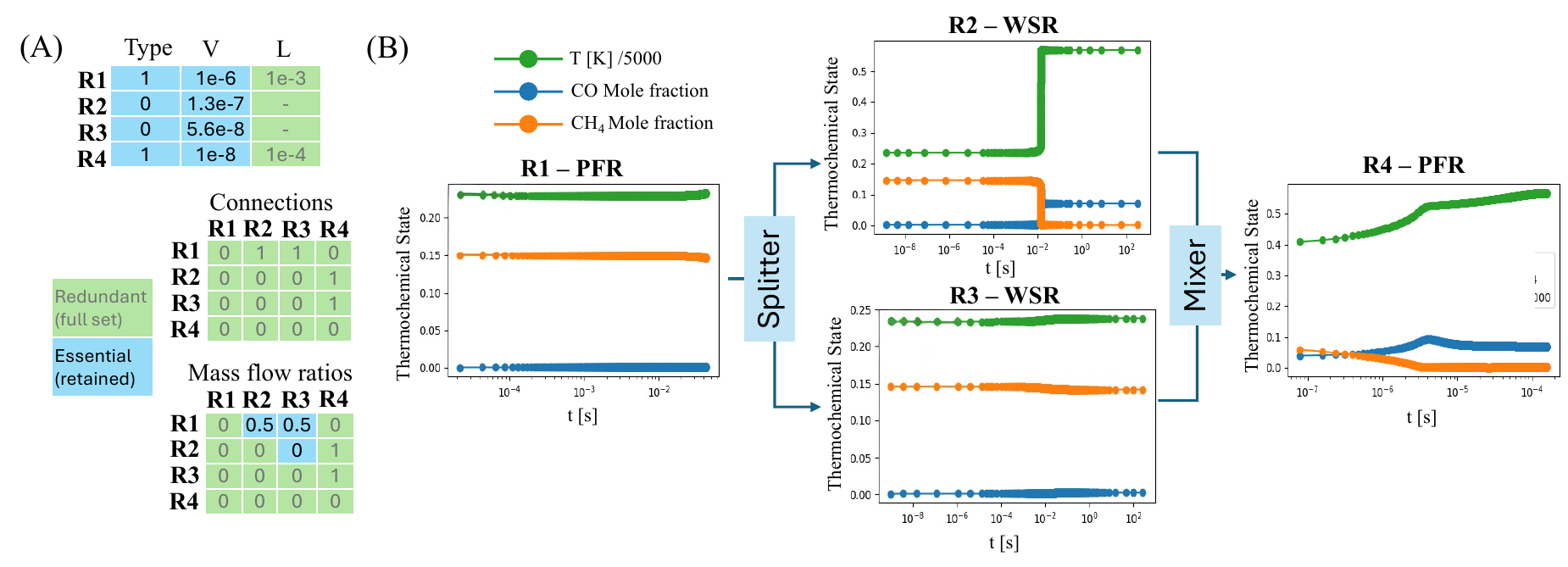}
	\caption{(A) Example encodings for a 4-reactor network. Type 0 is WSR, while type 1 is PFR. V and L are volume and length, respectively. Reduced set uses blue entries only (see Table \ref{tab:params}). (B) ERN simulation encoded by values in (A): main fuel (orange), CO pollutant (blue), normalized temperature (green).}
	\label{fig:encoding}
\end{figure*}

\subsection{Efficient design parameterization and reactor network design}\label{Sec:Param}
Both reactors require a few key quantities to be fully defined. A WSR is defined by just its volume, while a PFR requires a length and cross-sectional area. When constructing an ERN, the connections between reactors and mass flow rate ratios must also be defined. Finally, to run a simulation, the global mass flow rate and thermochemical input must also be provided, although these quantities are provided by the simulation target and are not design variables. 

For $n$ reactors, we define $3n$ intrinsic parameters: one to define the reactor type (0 for WSR or 1 for PFR), a second to define the PFR length or WSR volume, and a third to define the PFR area. For example, an input reactor R1's intrinsic encoding of $[1, 1.5\times10^{-3}, 1\times10^{-3}]$ defines a PFR of length $1.5\times10^{-3}$ m and area $1\times{}10^{-3}$ m$^2$ (see a visualization in Fig.~\ref{fig:encoding}). In these ideal simulations with no heat loss or conduction, however, we found that the PFR length-to-area ratio is actually unimportant: so we are able to eliminate the third index completely, leading to $2n$ parameters per reactor: the type, and its volume. The same R1 is now $[1, 1.5\times10^{-6}]$.

The extrinsic connections between reactors, meanwhile, are bulky and show potential for large reduction. The complete representation for a set of $n$ reactors would entail an $n$x$n$ matrix defining all connections, where the $i, j$ entry being $0$ indicates no connection between reactors $i$ and $j$, while $1$ indicates a connection; and a second $n$x$n$ matrix defining the relevant mass flow ratios, where some $i, j$ entry of $0.5$ indicates that half of the mass flux exiting reactor $i$ enters reactor $j$. Thus for a network with n reactors, the connective representation would require $2n+n^2+n^2=2n^2+2n$ parameters. We find this representation needlessly bulky for a handful of reasons:

\begin{itemize}
    \item The use of both a masking matrix and a mass ratio matrix appears redundant. Zero-valued entries in the mass ratio matrix should implicitly encode zero-valued entries in the masking matrix. 
    \item The diagonal is entirely unnecessary, as the products of a reactor do not flow back into the same reactor.
    \item Mass conservation dictates that the sum of each row in the mass flux matrix must sum to unity: exactly 100$\%$ of the mass leaving reactor $i$ must enter some combination of one or more downstream reactors, or be the final simulation output.
    \item While not necessarily globally true for all reactor models, we typically do not see global recirculation in the referenced works, making a reasonable simplification that downstream reactors do not feed into upstream reactors.
\end{itemize}
\begin{table}[!htbp]
\caption{Parameter count comparison between the full and engineered ERN parameterizations.}
\label{tab:params}
\begin{ruledtabular}
\begin{tabular}{lcc}
\textbf{Reactor Count} & \textbf{Full} & \textbf{Engineered} \\[2pt]
3 reactors & 27 & 7 \\
4 reactors & 44 & 11 \\
5 reactors & 65 & 16 \\
6 reactors & 90 & 22 \\
\end{tabular}
\end{ruledtabular}
\end{table}

Based on these, we instead represent the reactor system with the intrinsic $2n$ parameters as well as a single upper triangular matrix, excluding the diagonal and excluding the last column, of $(n^2-3n+2)/2$ parameters, for a  grand total of $(n^2+n+2)/2$ parameters. Full and reduced parameter counts are shown in Table \ref{tab:params} for a few representative sizes, while Fig. \ref{fig:encoding}(A) graphically depicts the parameter reduction. The last and penultimate reactors (R3 and R4 in a 4-reactor setup) require no mass flow encodings, as the penultimate must flow fully into the last due to no backwards flow, and the final reactor output is the simulation output. Finally, Fig. \ref{fig:encoding}(B) takes the same uninformed set of parameters depicted in Fig. \ref{fig:encoding}(A), generates the encoded reactor network, and solves it. Three tracked quantities are plotted (methane fuel, CO pollutant, and normalized temperature), while the complete solution includes 51 other major and minor species. R1 is seen to initiate the reactions in a PFR (near the 1-D inlet nozzle), with a slight increase in temperature and decrease in methane toward the output of the reactor. R2 and R3, two parallel WSRs to encode the recirculation zones and turbulent behavior, further react the output of R1 (note that, as per the chemical reactor network configuration, the output thermochemical state of R1 is exactly that of the R2 and R3 inputs). The larger-volume R2 has a larger residence time and runs the reaction largely to completion, while the smaller-volume R3 has a smaller residence time (due to the same mass flow rate through a smaller mixed volume), and only slightly reacts the system. Finally, the R2 and R3 outputs are mixed as the inlets to R4, the downstream PFR, where an initial curvature represents methane combustion, and when methane is exhausted a change in curvature represents further reaction involving CO and other minor species. These parameters were manually tuned to provide an interesting solution, but are not fitted to the simulation data. Sec. \ref{sec:ML} describes the ML pipeline used to optimize the ERN.

\subsection{Combustion dataset}\label{Sec:CFD}
\begin{figure}[!htbp]
    \hspace*{0.5cm}
    \includegraphics[width=1\linewidth]{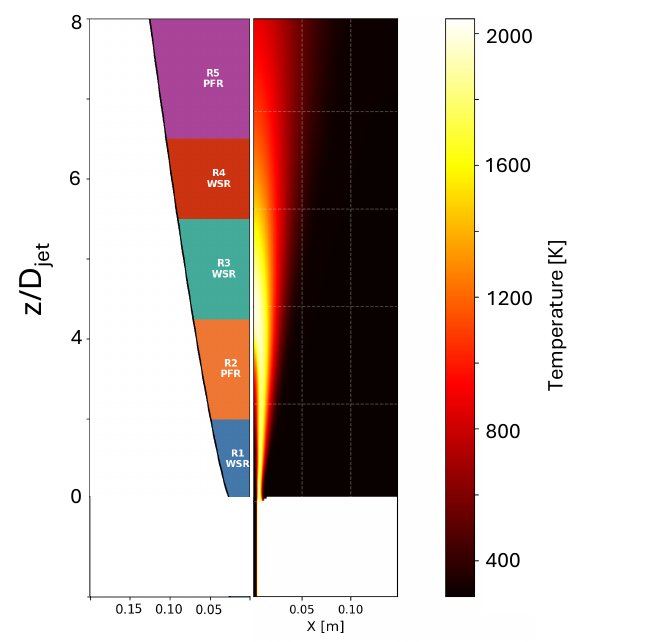}
    \caption{A visual of a simple ERN configuration (left) alongside the RANS simulation of the flame (right). $D_{jet}$ is the jet diameter of the flame.}
    \label{Fig:Sandia}
\end{figure}

For downstream testing of ERNs, we generated a dataset of computational fluid dynamic (CFD) simulations of the canonical Sandia-D flame configuration~\cite{BARLOW19981087} (Fig.~\ref{Fig:Sandia}). This flame consists of an inlet jet of CH$_4$/air issuing from a 7.2 mm nozzle, surrounded by a hot stoichiometric pilot and an outer coflow of air. The central jet is fuel-rich ($\phi\approx3.2$) and exits at $\sim$50 m/s, while the pilot supplies 1880 K burned products that stabilize the reaction zone. A slow air coflow ($\sim$0.9 m/s) envelopes the jet, producing the partially premixed turbulent diffusion structure.
Further flame details are in Ref.~\cite{BARLOW19981087}. 

We chose this flame because it provides a diverse range of non-premixed, mixing, and premixed zones where a combination of PFRs/WSRs may act as useful representations of chemistry. Other papers have also investigated representing Sandia-D using standard, non-ML ERN configurations such as those in Refs.~\cite{MonaghanEF2012} and \cite{MODARRES2025100893}, further confirming that ERN representations are possible, and providing baselines for us to compare ERN performance.

Our dataset consisted of 30 Reynolds averaged Navier-Stokes (RANS) simulations at five inlet pilot temperatures evenly spaced between $1600$ K and $2400$ K and six inlet CH$_4$ mole fractions evenly spaced between $0.1$ and $0.2$. The \texttt{reactingFoam} solver was used within the OpenFOAM open-sourced CFD framework.
We used detailed chemistry, with the GRI-3.0 mech with \textit{in-situ} adaptive tabulation (ISAT) for acceleration~\cite{GRImech3}. The total computational running time was 5m 30s per run on 6 cores resulting in a total of $\sim16.5$ CPU hours for all $30$ runs.

\subsection{Baseline tests} \label{sec:baseline}
We first attempted to apply the ERN framework described in Sec.~\ref{Sec:Param} to the flame configuration described in Sec.~\ref{Sec:CFD} using a reactor configuration based on intuition.
This initial test acts as a baseline to compare our ML-automated efforts.

To begin, we chose a simple serial configuration presented in Fig.~\ref{Fig:Sandia}B.
This ERN consisted of five steps: (1) an initial Mixing Zone (WSR) to represent the jet and pilot mixing with equilibrium ignition. Since the pilot represents vitiated flow, steps were taken to ensure the composition of this inflow region was chemically equilibrated to the appropriate mixture conditions. (2) Primary Reaction Zone (PFR). (3) Secondary reaction zone (WSR) for continued reaction. (4) CO burnout zone (WSR) with high temperature product-region oxidation. And (5) the final zone (PFR), where coflow mixing and far-field reactions occur.

We chose two metrics to assess the performance of the ERN: accuracy of reconstructing net CO emissions and the accuracy of peak temperature. These two metrics were chosen because CO is a key pollutant, and tracing its production from an engineering design configuration is an essential task for many combustion-based industries. 
Temperature is also important because it acts as a global combustion metric that ties directly to total fuel consumption and overall flame intensity. 

\begin{figure*}[!htbp]
    \includegraphics[width=0.9\linewidth]{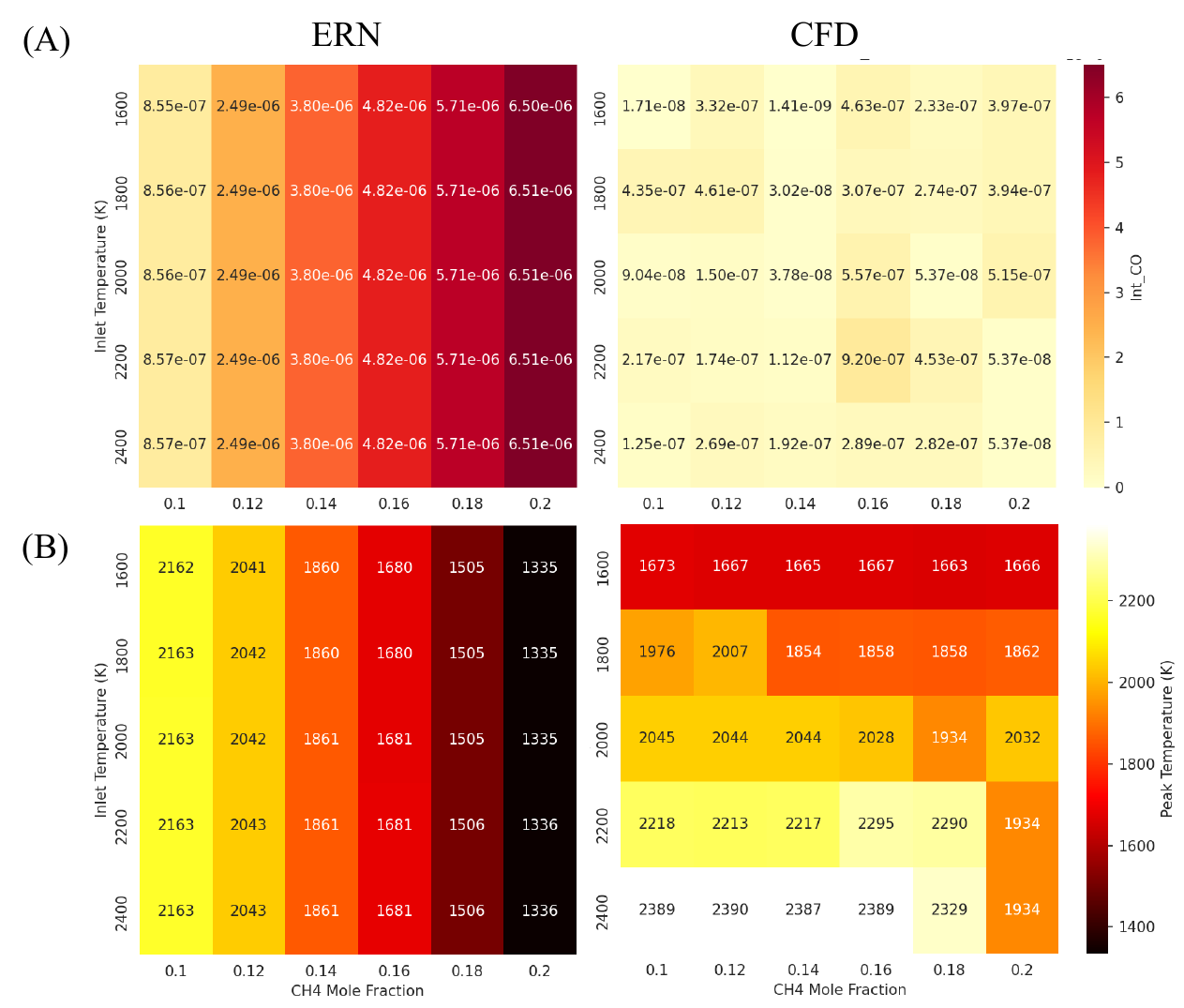}
    \caption{(A) CO emissions and (B) peak temperatures in the Sandia D flame as predicted from the manually-constructed ERN and CFD simulations.}
    \label{Fig:Performance}
\end{figure*}

Initial test results are presented in Fig.~\ref{Fig:Performance}. The computational runtime for the 30 Sandia-D configurations using the ERN was only 12.9 seconds on a single CPU core, resulting in a $\sim4600\times$ speedup over the RANS-CFD simulation (which themselves were accelerated $\sim$10x using tabulated chemistry).
However, the initial guess configuration performed poorly compared to RANS results. Reactor volumes were assigned as $V=\{5\times10^{-4},0.05,5\times10^{-4},5\times10^{-3},0.3\}$ m$^3$ for reactors one through five, respectively. These volumes must be tuned better to partition the overall volume accurately between the different reactor types. Furthermore, the count, layout, configuration, and mass flow fractions of these reactors must be better coordinated to emulate the combustion physics. For example, the mixing process included in the Sandia-D flame between the inlet jet and pilot streams is a complex process that cannot be modeled by either the WSR or PFR independently. A combination might be suitable, to stagger and better implement  mixing.

\section{ML-automated pipeline} \label{sec:ML}

\begin{figure*}
    \centering
    \includegraphics[width=0.8\linewidth]{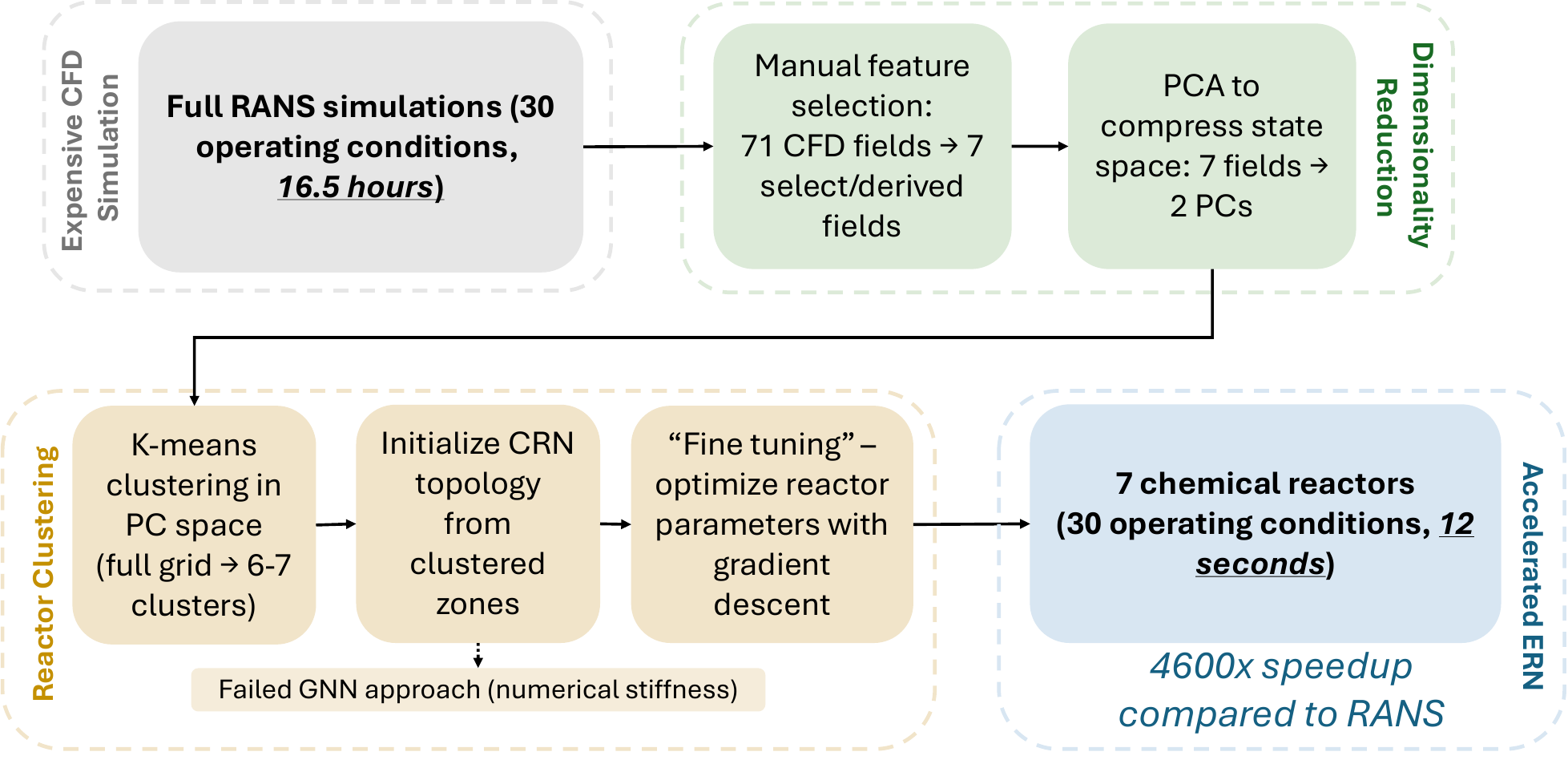}
    \caption{Overview of the complete ML pipeline used in this work.}
    \label{fig:overview}
\end{figure*}

The complete ML pipeline includes principal component analysis and k-means clustering to define an ERN initialization based on the real CFD data, then final tuning where we attempted graph-based methods before turning to gradient descent. An overview of the complete methodology is shown in Fig. \ref{fig:overview}.

\subsection{Principal component analysis}
Our first step to partitioning the flow domain into reactors was to identify the most optimal flame regions to delegate the reactor models. These models are typically highly concentrated around the main flame region, but identifying their boundaries is a challenge with the high-dimensional feature space and complex flow-field structure. Each grid cell in the combustion simulation consists of 71 features, including grid cell location, temperature, pressure, mole fraction of every species, and RANS-specific data including turbulence metrics. In machine learning, constructing a clustering algorithm that faithfully partitions is challenging with this many dimensions because all points start becoming equally distant and some dimensions with small physical relevance can distort clusters.  In combustion specifically, chemical species and thermodynamics are governed by conservation laws and chemistry, implying they don't vary independently and are highly correlated \cite{ji2019quantifying}. 

In order to compress this feature space and remove redundant information, we first reduce our dataset to seven features: $\theta=\{T, Z_{B},Y_{CH_4},Y_{CO},Y_{O_2},Y_{CO_2},Y_{H_2O}\}$, for temperature $T$, mass fraction $Y_k$ for species $k$, and Bilger mixture fraction $Z_{B}$, defined as
\begin{equation}
Z_{\mathrm{B}} =
\frac{
\left( 2 Z_C + \frac{1}{2} Z_H - Z_O \right)
-
\left( 2 Z_C + \frac{1}{2} Z_H - Z_O \right)_\mathrm{o}
}{
\left( 2 Z_C + \frac{1}{2} Z_H - Z_O \right)_\mathrm{f}
-
\left( 2 Z_C + \frac{1}{2} Z_H - Z_O \right)_\mathrm{o}
}
\end{equation}
for elemental mass fractions $Z_i$ for element $i$ evaluated locally and at the fuel ($f$) and oxidizer ($o$) inlets. 
We then apply principal component analysis (PCA) on these features, following a similar approach to Ref.~\cite{IHME2022101010}. 
This not only reduces the dimensionality of our system but also acts as a denoiser for spurious flow-field numerical perturbations and irrelevant features. Finally, the reduced feature space enhances the interpretability and computational efficiency of the downstream clustering algorithm.

\begin{figure*}
    \centering
    \includegraphics[width=1\linewidth]{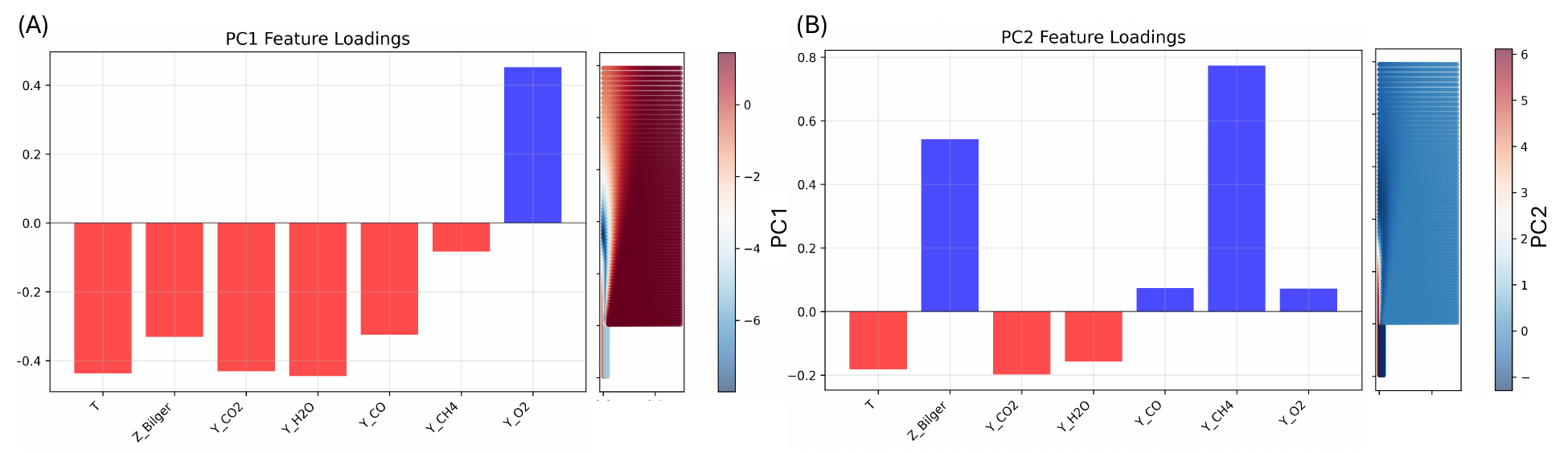}
    \caption{PC loadings among the seven input features and contours of PC in the flame for the (A) first and (B) second principal components.}
    \label{fig:PreProcessing}
\end{figure*}

This unsupervised algorithm enables the automated extraction of dominant modes of variation in thermochemical space, producing modes which align closely with concepts in the flame physics. Figure~\ref{fig:PreProcessing} presents the first two most dominant principal components of the Sandia-D flame configuration with $1800$K pilot inlet temperature and CH$_4$ inlet mole fraction of $0.10$. 
The first mode aligns with increasing O$_2$, and negative loadings on the products, fuel, and Z$_B$, suggesting PC1 distinguishes unreacted portions of the domain from the burning core. This aligns with the combustion progress/dilution axis of the Sandia-D flame.
PC2 has high CH$_4$ and Z$_{B}$ loading, indicating when PC2 is high, we are in the fuel-rich jet core and preheat region, while when PC2 is low, we are in the oxidizer and products region. This aligns with the mixing aspect of the flame.
In effect, PCA is able to distinguish the two axes commonly discussed by combustion experts but in an automatic, unsupervised fashion: the progress variable axis, which defines the extent of progression from reactants to products, and the mixture fraction axis, which defines the mixing extent between fuel and oxidizer.
These two modes were able to account for $>90\%$ of the variation in the flame, suggesting these two axes are highly dominant in representing the flame structure.

\subsection{K-Means clustering}
After conducting dimensionality reduction, the flame is partitioned using the unsupervised k-means clustering algorithm into regions. These regions provide a strong initialization structure for the reactor configuration. K-means has been applied before to flames, as in Ref.~\cite{IHME2022101010}, but applying PCA and then k-means for ERN creation has only been applied in a limited capacity~\cite{SAVARESE2023127945}.

\begin{figure*}
    \centering
    \includegraphics[width=1\linewidth]{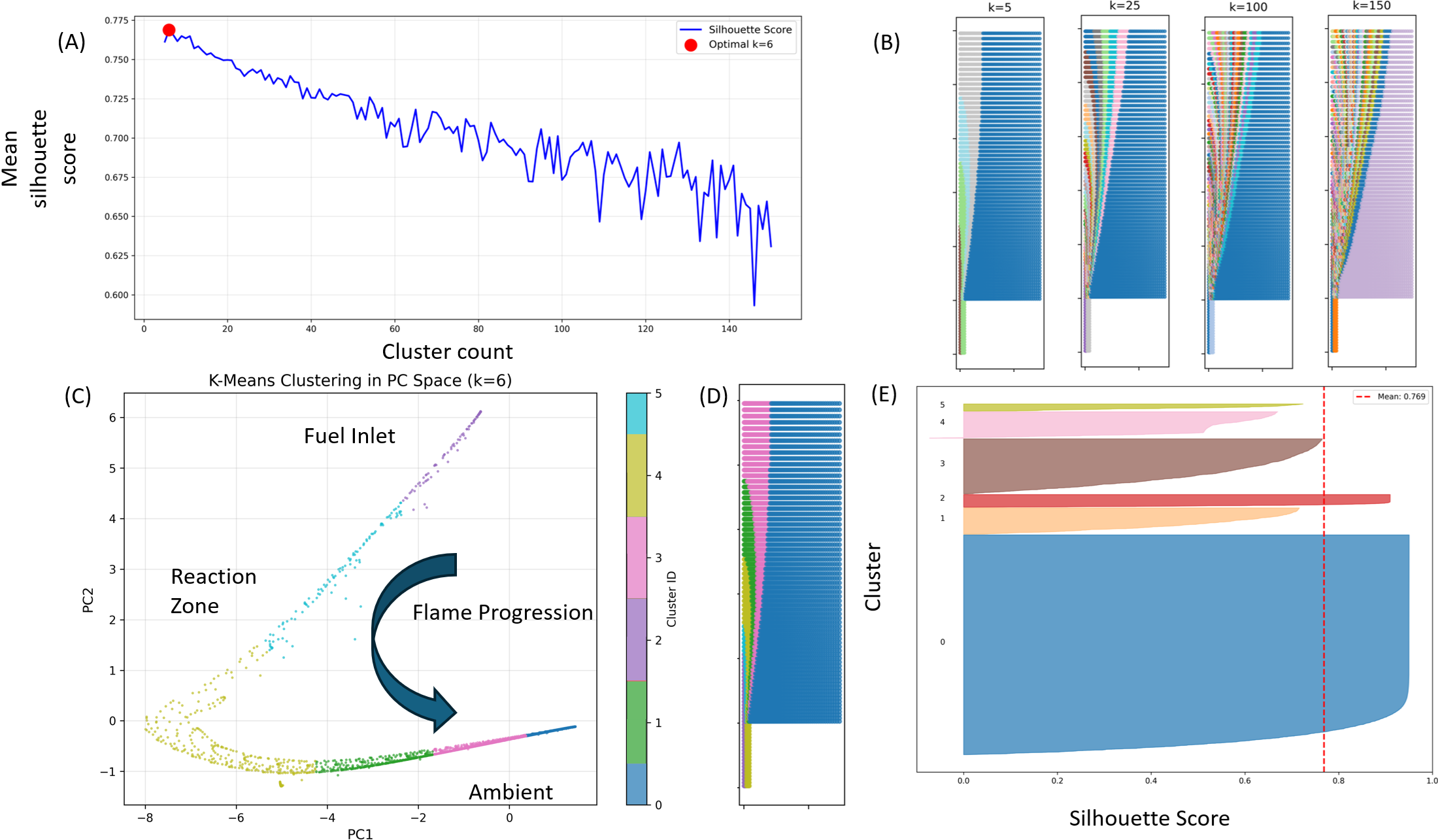}
    \caption{K-means clustering results. (A) Silhouette scores at varying cluster counts, (B) various clustering results at increasing k-counts. (C) K-means clustering regions in PC-space, (D) in physical space, and (E) silhouette scores for optimal $k=6$.}
    \label{fig:kmeans}
\end{figure*}

We apply k-means using the first two principal components and RANS simulation data corresponding to a pilot temperature $1800$ K and inlet jet CH$_4$ mass fraction of $0.1$ (one of 30 CFD profiles).
We varied the cluster count from 5 to 150 and determined the optimal value using the silhouette score. The silhouette scores for the range of k-counts and the silhouette plot for the most optimal k-count are presented in Fig.~\ref{fig:kmeans}A alongside several clustering results in~\ref{fig:kmeans}B. It was found that just six clusters provided the highest silhouette score.

Similar to PCA, the k-means clustering algorithm provides an interpretable, unsupervised latent representation of the flame configuration that corresponds to real flame physics. The latent and physical space representations of the k=6 clusters are presented in Fig.~\ref{fig:kmeans}C, D and E. The latent clusters partitioned the flame into inlet jet, inlet pilot, low-temperature ignition, high-temperature combustion, product regions, and ambient zones. Within the principal-component space, these are projected into a two-pronged shape, with the upper prong corresponding to the near-inlet conditions, and the lower prong corresponding to the outlet. Interestingly, there exists a sharp turnaround point where oxidation begins to occur and temperature is maximized. Here, PC1 reaches its minimum value (corresponding to the point where the least O$_2$ exists), and PC2 roughly flatlines as a function of axial coordinate.

It is important to recognize, however, that just because six clusters most optimally partition the flame in PC-space, it may not necessarily be the most optimal configuration of the reactors themselves, which each include 0-D or 1-D ODE solves over the complex and highly-stiff chemical rate equations, and include variations therein depending on their parameters and whether they are well-stirred reactors or plug flow reactors.
However, we postulate that clustering in this fashion provides an excellent initial-guess reactor configuration that can later be refined.

\subsection{Graph representation and cluster corrections}
After clustering, we organized our reactor configuration into a graph, with nodes as the reactors, and edges as mass-flow exchanges. The node attributes include volume and reactor type (WSR/PFR), and the edge attributes include mass flow (a quantity derived either from pressure controllers or a tunable parameter).

\begin{figure*}
    \centering
    \includegraphics[width=0.80\linewidth]{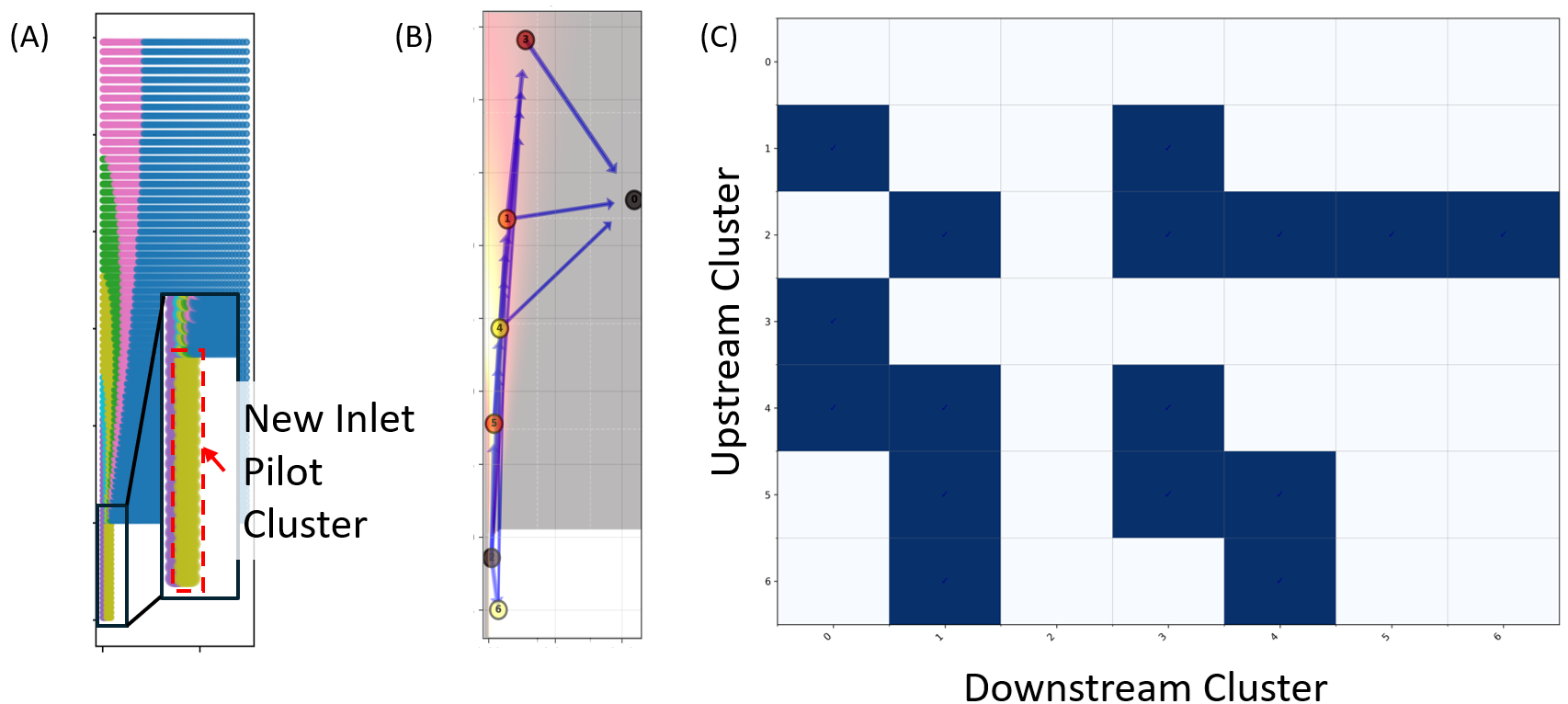}
    \caption{(A) New cluster at the pilot. (B) Graph configuration. (C) Reactor connectivity matrix (blue = connected).}
    \label{fig:graph}
\end{figure*}

As a preliminary step, we chose to include an additional region on top of k-means-identified clusters at the inlet, as we realized the cluster number 4 shown in yellow in Fig.~\ref{fig:kmeans} was encompassing both the inlet-pilot region as well as the core of the flame. Thus, we introduced a seventh cluster to encompass the inlet pilot region (see Fig.~\ref{fig:graph}A).

After introducing this 7th cluster, we were able to separate the pilot and fuel/air inlet streams and generated the reactor graph shown in Fig.~\ref{fig:graph}B with connectivity matrix Fig.~\ref{fig:graph}C. 
This graph is acyclic and directed, resulting in no mass flow recirculation as aligned with previous discussions (Sec. \ref{Sec:Param}).
The inlets from the fuel/air jet and pilot are assigned as constant mass flows to the local clusters, all calculated from the experimental Sandia-D configuration. The outlet regions are valves regulated by the upstream pressure. The mass flow is scaled down partially to reduce numerical stiffness.

Initial attempts were made to use a graph neural network (GNN)-based approach to classify reactors, graph edges, and regress their volumes and mass flows; however, it was quickly determined that the integration problem associated with the chemical ODEs and disparate reactor volumes was far too stiff for robust sensitivity-based neural network training. Ultimately, we elected to skip the graph-ML step for a simpler gradient-based approach described in a later section.

\subsection{Unrefined reactor configuration results}

As a first iteration, we used a WSR for each region with pressure-driven valves defining the mass flow exchanges between them. Reactor volumes are calculated geometrically by first gathering the area of each nodal region and then summing all the nodal regions assigned to that reactor. Nodal regions are calculated using the square-average distance of each node to its four nearest neighbors (K-NN approach). We tested all thirty Sandia-D configurations using the pressure-regulated mass flow conditions and uniform volumes assigned from clusters generated from just the $T=1800$ K and $Y_{CH_4}=0.1$ case.

\begin{figure*}
    \centering
    \includegraphics[width=1\linewidth]{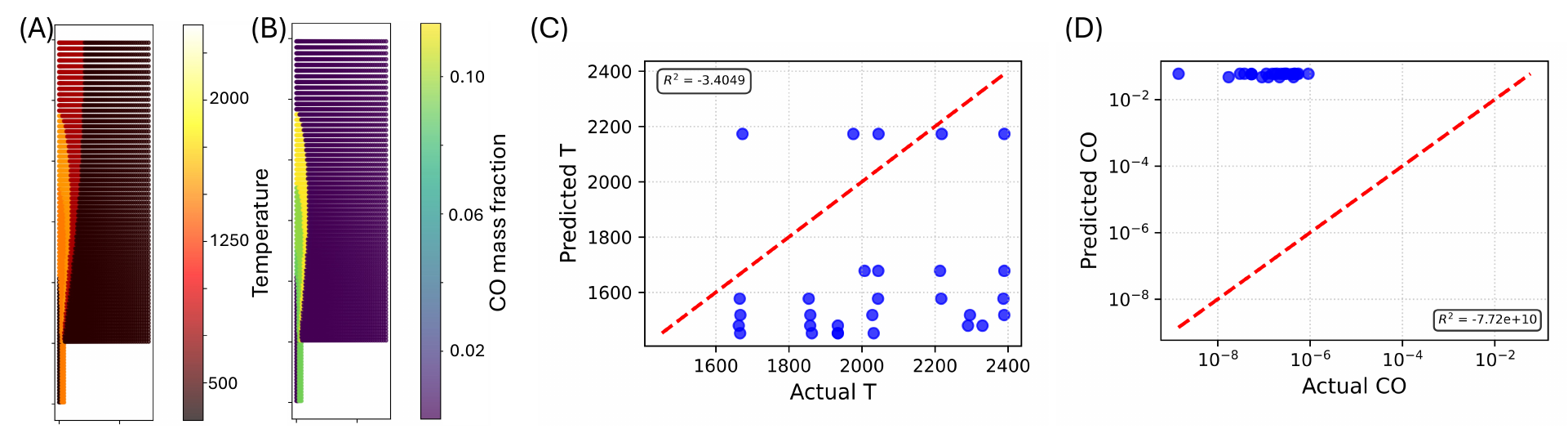}
    \caption{Initial results of unsupervised clustering graph approach to reactor initialization. (A-C) temperature contour and parity plot, respectively. (B-D) CO mole fraction contour and parity plot, respectively.}
    \label{fig:initial_results}
\end{figure*}

Results are presented in Fig.~\ref{fig:initial_results}. The per-ERN runtime was approximately one minute per case configuration, slowed significantly by the pressure-driven mass flow valves between each reactor. These valves made the solution procedure implicit, requiring iterations over the ERN to resolve local reactor thermodynamic/pressure-driven mass flow couplings. This slowdown was significant, reducing the CRN's speedup to only 10x over the CFD. The temperature and CO contours compared well between the CFD and the reactor networks, with each phase of the ERN evolving the chemistry in a realistic way. However, when compared on a parity plot and comparison Table~\ref{table:before_after_gd}, both temperature and CO emissions predictions were starkly different.

Our initial tests performed poorly against the CFD for several reasons. For one, the reactor volumes were based on the k-means clusters from just one of the cases, and these values fail to extrapolate well beyond that one case. This results in uneven and inaccurate chemical evolution within each of the 7 reactors. Additionally, while the reactor volumes and locations were based on the single CFD dataset, the connection between WSR/PFR physics and turbulent reacting flow physics is not as simple, and requires further optimization to accurately couple the two. In order to make the model more accurate across the range of inlet conditions, we required more elaborate training schemes and a broader training dataset.

\section{Gradient-Descent Enhancement}

For final tuning, we turned to gradient descent. While the CRN constructed via PCA and k-means clustering had poor performance and was tailored to a single dataset only, it offered the key benefit of being a reasonable starting point, which is nontrivial in such a chemical kinetic framework: most randomly initialized CRNs would never ignite or would immediately burn out in the first reactor due to the extreme numerical stiffness present in the governing equations \cite{ji2021stiff}. 

\begin{figure*}
    \centering
    \includegraphics[width=1.0\linewidth]{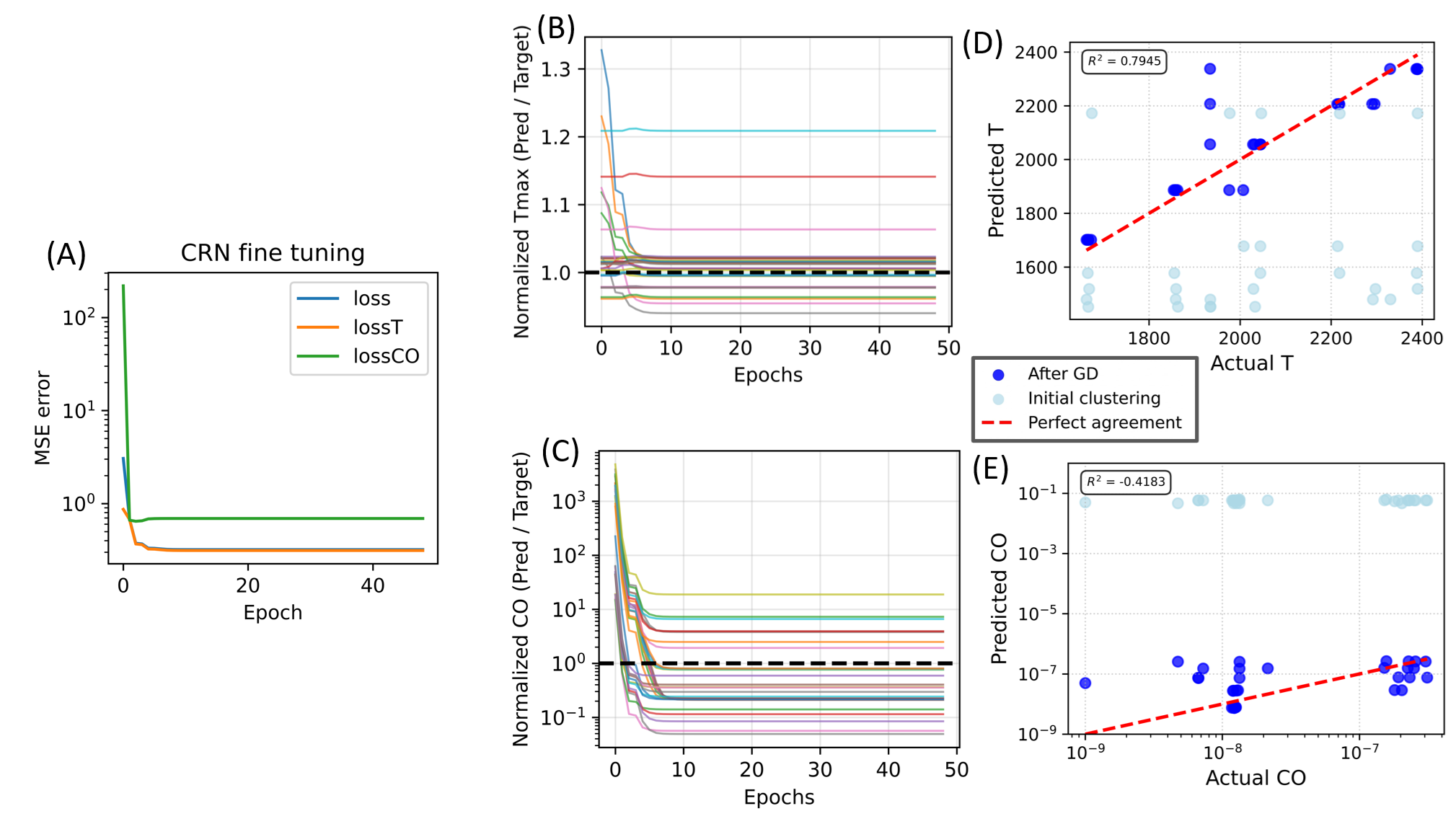}
    \caption{Final convergence with gradient descent in MSE (A), maximum temperature normalized to the actual value (B), and output CO normalized to the actual value (C). (D-E) Parity plots for maximum temperature and output CO, respectively. Note log-scale y-axis in CO plots due to fast chemistry and scale separation.}
    \label{fig:GD}
\end{figure*}

\begin{table}[!bp]
\caption{Coefficient of determination ($R^2$) for peak temperature and outlet CO predictions before and after gradient descent (GD) optimization.}
\label{table:before_after_gd}
\begin{ruledtabular}
\begin{tabular}{lcc}
\textbf{Optimization Stage} & \textbf{$T_{\max}$ $R^2$} & \textbf{CO $R^2$} \\[2pt]
Before GD & -3.4 & $-9.23 \times 10^{8}$ \\
After GD  & 0.7945 & -0.4183 \\
\end{tabular}
\end{ruledtabular}
\end{table}

We identified two general approaches for gradient computation: automatic differentiation and finite differences. 
Automatic differentiation would be the fastest and most accurate approach, but the Cantera implementations of the 0-D and 1-D reactors in Python are not written differentiably. To facilitate gradient-based optimization, we wrapped a Pytorch optimization framework around a Cantera finite difference solver. When gradients are requested, the solver is re-evaluated with a small perturbation applied to each input variable. The resulting changes in the outputs are then used to construct first-order forward-difference approximations to the gradients, according to
\begin{equation}
\frac{\partial y}{\partial x_i}(\bar{\mathbf{x}}) = \frac{y(\bar{\mathbf{x_i}'})-y(\bar{\mathbf{x}})}{\epsilon},
\end{equation}
where $y$ is the quantity of interest (here evaluated once for maximum temperature and once for CO output), $\bar{\mathbf{x}}$ is the original CRN encoding, and $\bar{\mathbf{x_i}'}$ is the encoding with the $i^{th}$ value perturbed by some small value $\epsilon$ (we used $\epsilon=10^{-9}$ for most cases), for example when perturbing the second value $\bar{\mathbf{x_2}'} = \bar{\mathbf{x}}+[0, 10^{-9}, 0, 0, ...]$. While not as accurate or efficient as automatic differentiation, this approach enabled a relatively quick implementation for preliminary testing. Computational costs of the FD approximations, which required new finite difference solutions for each degree of freedom in the model, were kept reasonable thanks to the efficient model parameterizations introduced in Sec. \ref{Sec:Param}. We additionally reverted to the explicit solution procedure described in Sec. \ref{sec:baseline}, in which inter-reactor mass flow rates are held fixed rather than obtained through an implicit pressure-coupling solve. This simplification results in a 6000× speedup for each ERN evaluation over the CFD, making the additional computational cost associated with finite-difference (FD) sampling tractable. Finally, we used a weighted MSE loss function to combine the maximum temperature and CO output contributions according to their relative mean values,

\begin{align}
\mathcal{L}(\mathbf{x}) = \frac{1}{2000} \sum_{i=1}^{30} MSE(T_{max, i}^{actual}-T_{max, i}^{pred}) \\+  \frac{1}{10^{-8}} \sum_{i=1}^{30} MSE(CO_{out, i}^{actual}-CO_{out, i}^{pred}). \nonumber
\end{align}

Reactor volumes and mass flow fractions were updated using FD-based GD. To demonstrate the versatility of the method, we switched from an all-WSR network to a network with two PFRs (reactors 3 and 5, for similar justification as in Sec. \ref{sec:baseline}). The convergence plot in Fig. \ref{fig:GD}(A) shows that the optimal solution was rapidly found. Thus, while the FD gradients were slow to compute (3-5 minutes per epoch, where each epoch had on the order of 500-1000 forward solves), only a handful of epochs were needed to rapidly converge the results, mitigating the overall cost. The cost of gradients was also mitigated by the efficient parameterization introduced in Sec. \ref{Sec:Param}. Fig. \ref{fig:GD}(B) shows the convergence of the maximum temperature, normalized to the true value. While 3 boundary conditions (of 30 total) appear to have not moved much during gradient descent, the remaining 27 cases were all well within 10$\%$ of the true value, with a rather dense cluster very close to the true value. The corresponding parity plot in Fig. \ref{fig:GD}(D) corroborates this, with 22/30 boundary conditions seen to lie very close to the y=x line. The post-GD R$^2$ score of 0.7945 was a significant improvement over the initial clustering guess R$^2$ score of -3.4.

Similar results are seen for the outlet CO in Figs. \ref{fig:GD}(C) and (E). Note that both of these plots have a logscale y-axis, to accommodate the multiscale CO behavior (while it is a key pollutant, it is also a minor species with extremely small and highly sensitive concentrations). Convergence here is notably not as strong as in the maximum temperatures, although we do remark that the MSE dropped significantly (Fig. \ref{fig:GD}(A)) and that the CO values did move significantly in the right direction (Fig. \ref{fig:GD}(C)). The converged R$^2$ score was -0.4183, which is below zero and suggests that simply using the mean value would have provided a better metric. However, we once again refer to the literature, where CO is known to be a highly challenging pollutant to model due to its fast chemistry, small scales, and interaction with turbulence, with even CFD models seeing high errors up to and around an order of magnitude \cite{klarmann2018numerical}. We therefore believe that even an R$^2$ score of -0.4183 is a decent starting point, especially in light of the starting point score of -9.23 $\times 10^8$, a much poorer result. In any case, the MSE loss for CO was seen to converge well in Fig. \ref{fig:GD}(A), and the qualitative log-scale accuracy of Fig. \ref{fig:GD}(E) additionally looks promising.

We finally close the loop by returning to Fig. \ref{fig:Modarres}, reproduced from the recent 2025 work of Ref. \cite{MODARRES2025100893} where a physics-based reactor network assignment algorithm was applied. Our metrics and reconstruction approach differ slightly from that work making direct comparison challenging, but we highlight that in a limited temperature profile comparison, our model appears to outperform this benchmark in maximum temperature reconstruction, especially in its low-reactor-count ERNs (for example, the 5 CSTR solution of Fig. \ref{fig:Modarres}, which misses most of the spatial evolution of the flame while also missing the maximum temperature by a full factor of two). This performance boost largely comes from the final GD tuning, while the coarser physics-based algorithm of \cite{MODARRES2025100893} did not incorporate a numerical optimization stage.

We conclude this section by once again highlighting how rapid the convergence of this reactor network was (Fig. \ref{fig:GD}(A)), which we found to be thanks to the highly physics-informed initial starting point that was provided by the PCA and k-means based initialization.

\section{Conclusions and Outlook\label{sec:Conclusions}}

We use machine learning to automatically reduce a computationally expensive combustion CFD simulation of the Sandia-D flame into relatively inexpensive ERN of 0-D and 1-D reactors. The ERN achieves strong accuracy compared to similar efforts in the literature and a 6000x speedup over the CFD solver, which itself already employs tabulated chemistry to accelerate the underlying kinetics..

We first applied PCA to reduce the high-dimensional thermochemical space to a two-dimensional latent space. The two principal component axes were able to account for over 90\% of the variation in the flame field properties, and each represented relevant field quantities including (1) the degree of flame progression from reactants to products with ambient diffusion, and (2) the mixing degree between reactants and products (\textit{i.e.}, Bilger mixture fraction).
The two principal components were then used in a k-means clustering step to break the flame into initial-guess reactor region for the ERN model. Despite testing up to $k=150$ clusters, it was found that only six clusters maximized the Silhouette score, and generated a physically interpretable breakup of the flame into physically distinguishable regions.
The clustering is similarly interpretable in the PC-space, producing a pronged shape with separation between the flame inlet, outlet, and combustion, demonstrating our latent-space representation of the flame is effective at differentiating flame regions. These clusters were then applied to a graph structure with nodes and edges corresponding to reactors and their mass flow exchanges, respectively. A seventh reactor was added near the inlet to capture the pilot region of the flame. Each node was initially assigned a well-stirred reactor model using volumes and pressure-regulated mass-flows evaluated from the CFD. This initial-guess reactor-model performed poorly at recreating peak temperature and CO emissions from the CFD, but did a good job of providing initial guess configurations for reactor refinement.

Final refinement leveraged gradient descent powered by a finite difference loop wrapped around the non-differentiable Cantera reactor code. Thanks to the strong initialization provided by the PCA + k-means clustering algorithm, the gradient descent converged rapidly in just a few epochs to a significantly improved result. The final R$^2$ score for the maximum temperature across all boundary conditions was 0.7945, while the much more challenging and scale-varying CO outlet saw a converged R$^2$ score of -0.4183 (negative, but a huge improvement over the PCA+k-means initialization of -9.23 $\times 10^8$).
 Limited comparison to recent physics-based ERN construction highlighted the high relative accuracy of our learned ERN's maximum temperature values, the majority of which were within a few percentage points of the true value compared to recent physics-based automated ERN generation, where such values for similarly-sized reactors were off by a full factor of two. We additionally highlight the promise of our tool in custom performance optimization: while the clustering algorithm provides a generally sound initialization, the gradient descent can be easily modified for other target quantities, like total heat release or fuel consumption efficiency. With such changes, end users can automatically create ERNs for their own specific performance needs.
 
 We finally highlight key areas for future refinement. We acknowledge that our attempts at a graph-based learning approach were not fruitful due to the high numerical stiffness of the problem, but propose that further work in this area could be of benefit. 
 Additionally, as in previous work~\cite{MODARRES2025100893}, it could be worthwhile to study reactor networks of different sizes (where our final GD convergence used a fixed, 7-reactor size). We chose our reactor count based the Silhouette score, which is a useful metric for deciding k-count in k-means clustering, but may not adequately optimize the accuracy of the downstream ERN solves. Likewise, two PCs may account for high variance, but additional PCs may have an influence on downstream network accuracy. Additionally, while our CFD data did not track NOx pollutants, these are also a significant quantity of interest for many combustion scientists. In fact, the GD approach is generally easily customizable for different outputs: further study of multi-objective Pareto fronts and the changes in shapes and mass flows of ERNs as the relative accuracies to different targets change could be useful. With more computational time and implementation effort, one could even introduce full temperature and CO profile accuracies, rather than maximum or output scalar value metrics.



\section*{Acknowledgements}
The work is supported by the National Science Foundation (NSF) under Grant No. CBET-2143625. BCK is partially supported by the NSF Graduate Research Fellowship under Grant No. 1745302.
NJT is partially supported by GE Vernova under the GE Vernova X MIT fellowship.
The authors also gratefully acknowledge support from the Kavli Foundation, Carbon Hub, and ExxonMobil.
This work is based on a final project for MIT course 2.156.


\bibliography{2156.bib}


\newpage

\small
\baselineskip 9pt


\end{document}